\documentclass[aps,prl,superscriptaddress,showpacs,preprint]{revtex4-1}
\usepackage{graphicx}

\begin{document}



\title{Silicon-carbon bond inversions driven by 60 keV electrons in graphene}

\author{Toma Susi}
\email{toma.susi@univie.ac.at or toma.susi@iki.fi}
\affiliation{University of Vienna, Faculty of Physics, Boltzmanngasse 5, A-1090 Vienna, Austria}

\author{Jani Kotakoski}
\affiliation{University of Vienna, Faculty of Physics, Boltzmanngasse 5, A-1090 Vienna, Austria}
\affiliation{University of Helsinki, Department of Physics, P.O. Box 43, FI-00014 Helsinki, Finland}

\author{Demie Kepaptsoglou}
\affiliation{SuperSTEM Laboratory, STFC Daresbury Campus, Daresbury WA4 4AD, United Kingdom}

\author{Clemens Mangler}
\affiliation{University of Vienna, Faculty of Physics, Boltzmanngasse 5, A-1090 Vienna, Austria}

\author{Tracy C. Lovejoy}
\author{Ondrej L. Krivanek}
\affiliation{Nion Co., 1102 8th Street, Kirkland, Washington 98033, USA}

\author{Recep Zan}
\affiliation{School of Materials, The University of Manchester, Manchester M13 9PL, United Kingdom}
\altaffiliation{Present addresses: R.Z. -- Department of Physics, Faculty of Arts and Sciences, Ni$\textrm{\u{g}}$de University, 51000 Ni$\textrm{\u{g}}$de, Turkey; U.B. -- Department of Physics and Energy, University of Limerick, Limerick, Ireland}
\author{Ursel Bangert}
\affiliation{School of Materials, The University of Manchester, Manchester M13 9PL, United Kingdom}

\author{Paola Ayala}
\author{Jannik C. Meyer}
\affiliation{University of Vienna, Faculty of Physics, Boltzmanngasse 5, A-1090 Vienna, Austria}

\author{Quentin Ramasse}
\affiliation{SuperSTEM Laboratory, STFC Daresbury Campus, Daresbury WA4 4AD, United Kingdom}

\date{\today}

\begin{abstract}
We demonstrate that 60 keV electron irradiation drives the
diffusion of threefold coordinated Si dopants in graphene by one lattice site at a
time. First principles simulations reveal that each step is caused by an
electron impact on a C atom next to the dopant. Although the atomic motion
happens below our experimental time resolution, stochastic analysis of 38
such lattice jumps reveals a probability for their occurrence in a good
agreement with the simulations. Conversions from three- to fourfold coordinated dopant
structures and the subsequent reverse process are significantly less likely than the
direct bond inversion. Our results thus provide a model of non-destructive and
atomically precise structural modification and detection for two-dimensional
materials.
\end{abstract}


\maketitle

Recent breakthrough developments in imaging and spectroscopy in (scanning)
transmission electron microscopy [(S)TEM] have enabled the study of structural
modifications that occur very literally at the atomic scale. Due to their low dimensionality, 
materials such as carbon nanotubes, and especially graphene, have proven ideal for these investigations~\cite{Meyer08NL,Suenaga10N,Zhou12PRL,Ramasse13NL,Bangert13NL,Nicholls13AN}.
At the same time, (S)TEM instruments can also be turned into nano-sculpting
tools: for example, graphene ribbons with specific geometries~\cite{Xu13AN}, or
perforations of controlled sizes~\cite{Ramasse12AN,OHern14NL}, can be
fabricated via adjustments of the local chemistry, electron beam
energy and density. Heteroatom doping is another way to tailor the properties 
of graphene~\cite{Terrones12RPP}, which is otherwise ill suited for many 
applications due to its lack of an electronic band gap~\cite{CastroNeto09RMP}. 
Exchanging some of the carbon atoms by boron or nitrogen can result
in an opening of the gap~\cite{Usachov11NL,Tang12AN}, while localised
enhancements of plasmon resonances can be created around single silicon
substitutions, which then act as atomic antennae~\cite{Zhou12NN}. 
The ability to directly observe the effect of single dopants is thus of
the utmost importance in the further development of nano-engineering. 

However, doping changes the effect an electron beam has on the atomic 
structure of graphene, as we have recently shown for nitrogen 
substitutions~\cite{Susi12AN}. In that case, the slightly higher
mass of nitrogen as compared to carbon leads to an increased likelihood to knock out the
carbon atoms next to the dopant than the dopant itself or C in pristine areas. 
We also predicted that damage would be negligible at a primary
beam energy of 60~keV, which was recently confirmed by atomic resolution
imaging and electron energy loss spectroscopy (EELS)~\cite{Nicholls13AN,Bangert13NL}.
Recent studies established that silicon dopants---significantly heavier and larger
in covalent radius than either carbon or nitrogen---can bond in two distinct ways within the
lattice: a non-planar, threefold coordinated configuration (denoted Si-C$_{3}$)
where the Si atom replaces a single carbon atom, binding to three neighbouring
C and buckles out of the plane; and a planar, fourfold coordinated
configuration (Si-C$_{4}$) where the Si atom is bonded to four C atoms and
occupies a divacancy in the lattice~\cite{Zhou12PRL,Ramasse13NL}. Although
beam damage was occasionally observed, apart from a study on the dynamics of
Si$_{6}$ clusters in a graphene pore~\cite{Lee13NC}, the effects of electron
irradiation have not been reported in detail.

In this Letter, we show that structural changes in silicon-doped graphene
(Si-graphene) drastically differ from those in nitrogen-doped graphene. For
Si-graphene, they predominantly take the form of a random walk by the Si atoms
through the lattice, with no other changes in the structure. Through first
principles molecular dynamics simulations, we show that each step is a result
of an electron impact on one of the C atoms neighbouring the Si, and how 
the non-planarity of the structure plays a crucial role. The
probability calculated for this process agrees well with an estimate obtained
through a stochastic analysis of the experimental data. We further discuss the
few observed events that lead to the conversion of Si-C$_{3}$ sites into
Si-C$_{4}$ ones, and show that they are well accounted for as knock-on
damage. The significantly greater probability of the non-destructive reorganisation 
coupled with its directionality should allow the motion of the Si atoms to be 
controlled with atomic precision.

\begin{figure}
\includegraphics[]{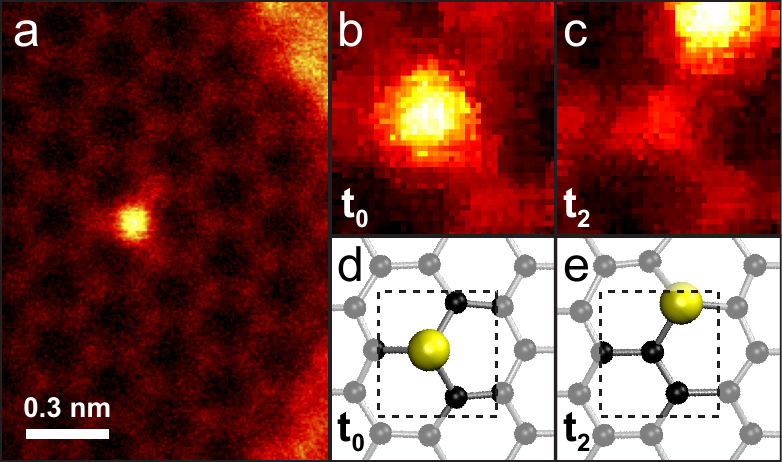}
\caption{(Color online) (a) An area of graphene where the bright contrast originates from a
single Si-C$_{3}$ dopant (high angle annular dark field (HAADF) detector). (b) A closer view of a Si site under continuous 60 keV electron irradiation in a reduced scan window at time $t_{0}$ (binned over six exposures of 88 ms, medium angle annular dark field (MAADF) detector). (c) The same area one binned frame (0.5 s)
later at time $t_{2}$, where the Si atom is observed to have moved by one
lattice site (MAADF). (d,e) Structural models where the areas
shown in (b,c) are outlined by the dashed lines.\label{Fig1}}
\end{figure}

Our graphene samples were synthesised by chemical vapour deposition~\cite{Li09Science}, and typically contain a low concentration of silicon incorporated as individual dopants in the lattice~\cite{Zhou12PRL,Ramasse13NL,Lovejoy12APL}. We observed the samples using a Nion UltraSTEM\texttrademark 100 electron microscope
equipped with a cold field emission gun operated at a 60~keV primary beam energy in near-ultrahigh
vacuum (2$\times10^{-7}$ Pa)~\cite{Ramasse13NL,SUPPL}. The Si atoms
can be directly identified within STEM images by their brighter contrast with
respect to the graphene lattice C atoms due to their higher atomic number~\cite{Krivanek10N}. The identification has also been verified by atomically resolved EELS~\cite{Zhou12PRL,Ramasse13NL}.

In Fig.~\ref{Fig1}, we illustrate the predominant beam-induced process, 
whereby the Si atom is seen to move from one lattice site
to the next during continuous imaging with an estimated dose rate of 2.2$\times10^{7}$~e$^{-}$/\AA$^{2}$s. Since atomic motion happens
at sub-picosecond timescales, electron microscopy can only capture static
snapshots of structures that are in effect fully relaxed. Although processes
happening below our experimental time resolution of 88 ms cannot be ruled
out, our observations give us a high degree of confidence 
that we capture the relevant dynamics. The observed events are
non-destructive reorganisations of the structure, similar to the process of 
Stone-Wales transformations that were earlier shown to be due to sub-threshold 
electron impacts~\cite{Kotakoski11PRB}, rather than thermally driven bond rotations~\cite{Li05PRB}.
The process is also not limited to perfect hexagonal arrangements (see the Supplemental Material~\cite{SUPPL}).

\begin{figure*}[t]
\includegraphics[width=0.75\linewidth]{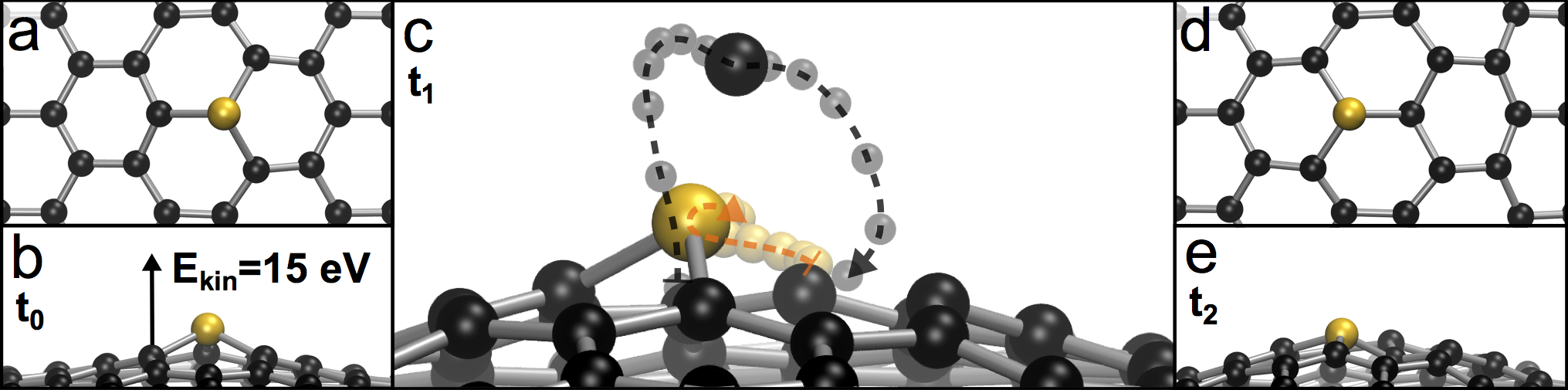}
\caption{(Color online) Molecular dynamics simulation of an electron impact delivering 
15~eV to a C atom neighbouring Si-C$_3$ in graphene. (a) Top view of the
starting configuration at time $t_{0}$. (b) Side view at $t_{0}$, with the kinetic energy 
indicated on the impacted atom. (c) A snapshot at $t_{1}$ ($\sim$700 time
steps into the simulation), with the entire trajectories of the ejected C atom
and the Si atom marked by semi-transparent balls and dotted lines. (d) Top view
near the end of the simulation at time $t_{2}$ after $\sim$1400 time steps. 
(e) Snapshot at time $t_{2}$. Note that although the atomic motion has not ceased by 
this point, no further changes in the atomic configuration follow.\label{FigSIM}}
\end{figure*}

To gain insight into the dynamics of such processes at the atomic scale, we
used density functional theory molecular dynamics (DFT/MD) calculations as
described in more detail in Refs.~\cite{Kotakoski11PRB, Susi12AN,Kotakoski12AN}. Although DFT describes the electronic ground state, the atomic dynamics take place over tens to hundreds of femtoseconds, whereas the relevant electron dynamics occur on sub-fs timescales in a metallic system \cite{Egerton13UM}. Ionisation effects were also explicitly ruled out by experiments with $^{12}$C and $^{13}$C graphene~\cite{Meyer12PRL}. Thus the ground state approximation is valid to a good degree of accuracy. In high-energy irradiation, the displacement threshold $T_{d}$ is defined as the minimum kinetic energy required by an atom to be removed from its position in a material. We estimated it by increasing the starting kinetic energy of a target atom until it escaped the structure during the course of an MD simulation. For atomic rearrangements, the same procedure was used to establish threshold limits for a particular reorganisation.

The calculations were performed using the grid-based 
projector-augmented wave code (GPAW~\cite{Mortensen05PRB,Enkovaara2010}; for
details on the computational parameters see Ref. ~\cite{SUPPL}). 
To speed up the calculations, we used a double-zeta linear combination 
of atomic orbitals (LCAO) basis. However, we 
directly compared the calculated knock-on thresholds for C in pristine graphene 
and the Si atom in Si-graphene with our earlier methodology~\cite{Kresse96CMS,Kresse96PRBa}, 
and established agreement within our computational
accuracy. Furthermore, we double-checked the Si jump threshold
by full accuracy finite-differences calculations.

The non-planar Si-C$_3$ configuration can have two distinct positions with 
respect to the beam direction: the Si atom either protrudes `below' (in the 
instrument geometry) the plane towards the incoming beam, or `above' it along 
the direction of the beam. These present different $T_d$ and cannot be experimentally distinguished from a 2-dimensional projection of the lattice recorded with a normally incident beam. However, we suspected that the threshold for flipping a Si atom from `below' to `above' should not be very high even for 60~keV electrons, as no bonds need to be
broken for the transformation. Indeed, a nudged elastic band~\cite{Henkelman2000JCP} 
calculation yields a barrier of only ca. 1.1~eV for this process (see Ref.~\cite{SUPPL}). Thus all Si configurations are effectively `above' the graphene plane under observation, and 
the beam-induced alignment is expected to be thermally stable.

Our simulations yield a $T_d$ of approximately 13.25~eV for the Si atom bonded in the Si-C$_3$ configuration. However, due to the large mass of Si (28~amu), the probability for
60~keV electrons to transfer this much energy to the dopant is low, resulting
in a cross section of 6.6$\times10^{-7}$ barn when out-of-plane lattice vibrations 
with a Debye temperature of 1287~K are taken into account~\cite{Meyer12PRL}. 
This agrees well with the observation that Si atoms are rarely lost~\cite{Zhou12PRL, Ramasse13NL, Lee13NC}. The knock-on threshold for one of the three C neighbours to Si is higher, about 16.875~eV, but leads to a much larger cross section ($\sim 0.022$~barn) due to the lower mass of C (unless otherwise stated, we refer to $^{12}$C). This process leads to a conversion
Si-C$_3$ $\rightarrow$ Si-C$_4$, which can occur at high irradiation doses.

More interesting are impacts below $T_d$, which
nevertheless lead to local changes in the structure. For example,
energies between 15.0 and 16.25~eV result in the C atom being ejected from the
lattice, but with a trajectory curving first slightly away and then towards the Si due to their mutual
interaction. The Si simultaneously relaxes towards
the vacated lattice site, as illustrated for the 15~eV-case in
Fig.~\ref{FigSIM} (a movie is available through Ref. ~\cite{SUPPL}).  
Between 16.5 to 16.75~eV the C atom almost escapes, but is
drawn back by the attractive interaction with the Si atom to land on top of the lattice 
on the side opposite to its starting position, while the
remaining structure assumes the Si-C$_4$ configuration. At 14.5~eV, the ejected
C is left as an adatom directly to the side of the Si, but at 14.75~eV, it
bounces off the Si atom on its downward trajectory and lands as an adatom on
the opposite side. For energies below 14.5~eV, no change in the structure is obtained.

Because Si-C$_3$ is energetically favoured over Si-C$_4$ (by ca.  1.02~eV~\cite{SUPPL}), it
is likely that all configurations where the C adatom remains very close will relax back into the Si-C$_3$ configuration (whether this results in an apparent jump event likely depends on whether 
C landed on the opposite side). We thus take 14.625~eV to be the lower threshold for this process, corresponding to a cross section of 0.494~barn. To estimate the cross section for moving the Si dopant, we subtract the cross section corresponding to $T_d$ (as the largest cross section is a sum of the cross sections of all possible outcomes), i.e., $0.494-0.022=0.472$~barn. If we instead assume no recombination of C adatoms with the Si-C$_4$ site, the cross section estimate is reduced to 0.316 barn. (These would be 0.130 and 0.084~barn for $^{13}$C, respectively.) All of the reported results correspond to displacements in a direction perpendicular to the graphene plane, and the curved trajectory is a result of the silicon-carbon interaction.

To double-check the calculated value, we carried out a few computationally demanding 
simulations using the default GPAW finite differences (FD) mode. 
The FD calculations gave a slightly lower knock-on threshold of 16.625 eV (0.032 barn), 
but also a lower flip threshold of
14.375 eV (0.666 barn); the jump cross section would thus be $0.666-0.032 =
0.634$~barn~\cite{note1}.

\begin{figure}
\includegraphics[]{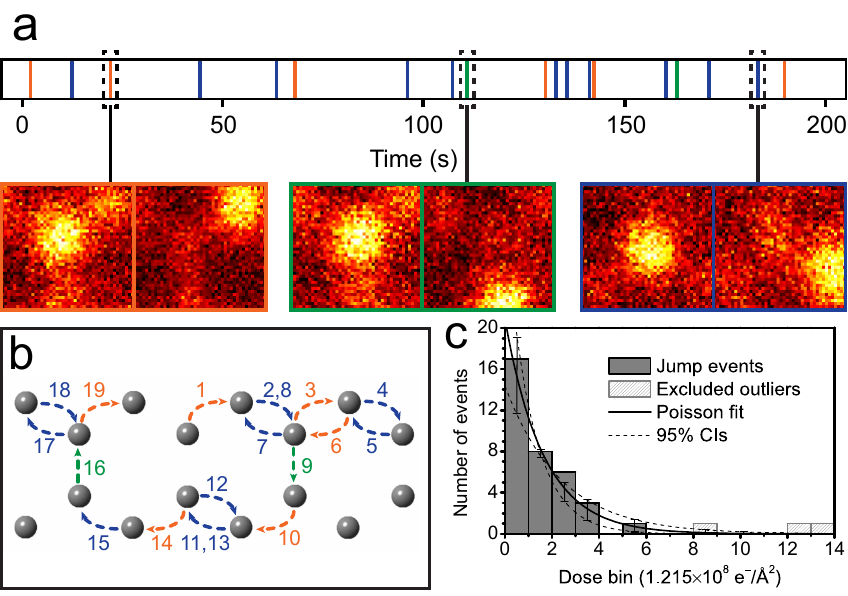}
\caption{(Color online) a) An experimental time series of 19 consecutive
jumps. Vertical lines mark the observed times of the jumps from
the beginning of the experiment. Each event has a colour corresponding to one
of the three inequivalent bonds in the lattice over which the jump occurred.
b) A schematic illustration of the directions of the jumps in the lattice. Insets: micrographs showing the third, ninth and 18th jump events (MAADF detector, each binned from six consecutive 88
ms frames). c) A histogram of the observed event doses with the optimised binwidth~\cite{SUPPL}. The solid line is an exponential with the fitted Poisson mean. The excluded outliers are shown in light grey (see text).\label{FigSTAT}}
\end{figure}

In our experimental data, we found 38 cases where a clear Si jump was observed,
with one continuous time series containing 19 consecutive jumps of the same Si
atom (Figure~\ref{FigSTAT}). The event doses (ie., the irradiation doses
between structural rearrangements) determined for the centre of the scan area
ranged from 0.24 to 20.01$\times10^{8}$~e$^{-}$/\AA$^{2}$. However, as the scan 
frame was centred on the Si atom, the doses on the neighbouring carbon atoms varied 
from scan to scan, which was taken into account via a Monte Carlo integration. 
If we assume that the data are stochastic, the waiting times (or, equivalently, the doses) 
should be Poisson-distributed. Thus the expected value of a Poisson distribution fitted 
to the data can be used to estimate a cross section for the process (see Ref.~\cite{SUPPL} for details).

We obtained an expectation value of 2.57$\times10^{8}$~e$^{-}$/\AA$^{2}$ for
the event dose, with a 95\% confidence interval (CI$_{95\%}$) of [1.59,
4.09]$\times10^{8}$ e$^{-}$/\AA$^{2}$~\cite{Khamkong2012}. 
This yields an interaction cross section of
0.389~barn (CI$_{95\%}$ [0.244, 0.629]~barn). However, further comparison
between the data and the obtained Poisson distribution reveals that three events
with the highest doses have a probability lower than $10^{-3}$ to be a
result of the same process. We believe that some of these can be due to $^{13}$C 
atoms incorporated into the graphene lattice (1.1\% of all C atoms 
can be expected to be $^{13}$C), or due to a combination of either a jump 
and an immediate reverse jump or a displacement and refill by an adatom. 
Reanalysis of the data without the outliers results in a revised expectation value of 1.63$\times10^{8}$~e$^{-}$/\AA$^{2}$ (CI$_{95\%}$ [1.12, 2.37]$\times10^{8}$ e$^{-}$/\AA$^{2}$), and 
a corresponding interaction cross section of 0.613~barn (CI$_{95\%}$ [0.423, 0.893]~barn)~\cite{note2}.

The experimental value is in remarkably good agreement with the FD calculation (0.634~barn), and within its CI$_{95\%}$ also with the range of [0.316, 0.472]~barn estimated from the more extensive LCAO simulations. Figure~\ref{FigSTAT}c shows a histogram of the event doses, which---as expected for a Poisson process---are found to be well described by an exponential with the fitted Poisson mean (apart from an excess of one event in the third bin, and a deficit of one event in the fifth).

\begin{figure}
\includegraphics[]{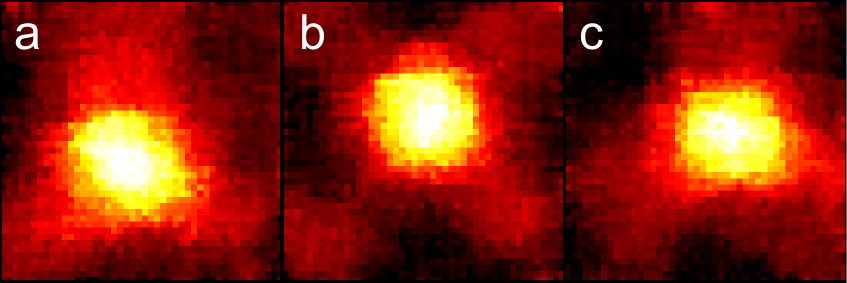}
\caption{(Color online) An example of Si-C$_3$ converting into Si-C$_4$ and 
back under 60 keV electron irradiation (HAADF). 
Frames processed with a Kuwahara noise-reduction filter (5$\times$5
kernel), each an average of six 156 ms frames. The dose rate 
was estimated to be 2.8$\times10^{7}$~e$^{-}$/\AA$^{2}$s. (a) The starting
configuration showing the Si atom bonded to three neighbours. (b) The
same area after three binned frames (2.8 s). A single carbon atom has been removed from
the site up from the Si in the image, and the Si has relaxed into fourfold
coordination with four carbon neighbours. (c) In the final
configuration, the site has reconverted to threefold coordination (after a
further three binned frames).\label{tetra}}
\end{figure}

Our data also contains eight events where Si-C$_3$ is transformed into
Si-C$_4$ (see Fig.~\ref{tetra}). Taking into account the total experimental dose on
trivalent sites (9.71$\times10^{9}$ e$^{-}$/\AA$^{2}$),
we get an estimated cross section of 0.08~barn for this process, in good
agreement with the calculated value of 0.066~barn~\cite{Note3}. Interestingly, the knock-on
threshold for the four C neighbours of the Si-C$_4$ was calculated to be
17.125~eV, i.e., slightly higher than that for the threefold site, suggesting that
Si-C$_4$ is more stable towards knock-on damage than Si-C$_3$. Indeed,
a fourfold site was only once observed to damage further by the loss of atoms.

Each of the eight Si-C$_4$ sites converted back into
Si-C$_3$ with the addition of a carbon atom (see Fig.~\ref{tetra}), presumably
by adatom diffusion~\cite{Zan2012,Gan08NJP} and subsequent recombination into the more
stable configuration. Under the non-equilibrium conditions of our experiment,
the average lifetime of the Si-C$_4$ configuration was ca. 70.0~s before
recombination. To understand this process, we performed additional structural
relaxation simulations of Si-C$_4$ sites with a single C adatom initially bonded to C-C bridge sites
1--4 bonds away from the Si. We found the total energy of the system with the C adatom 
at the closest site was $\sim$2.3 eV lower than when the C was three or four bonds
away (see Ref.~\cite{SUPPL} for details). This suggests that there is an attractive force drawing in mobile adatoms into the Si-C$_4$, possibly contributing to driving the observed recombinations. 

Although our analysis relied on the stochastic nature of 
events when all the area around the Si dopants was irradiated, it
should be stressed that due to the ca. $1$~\AA~beam diameter in modern 60~keV
aberration-corrected instruments, it is possible to restrict the
irradiation to a chosen carbon atom. For example, we estimate that a
1.1~\AA~beam with a Gaussian profile, centred on a selected C neighbour of the
dopant, would deposit $<0.3\%$ of the irradiation dose on the other two C
neighbours. Thus the motion of silicon atoms in the lattice can in
principle be controlled with atomic precision. To explore this idea, we calculated the total energies of systems with two Si atoms separated by 1--4 lattice sites, and found that the energies were lower for closer separations. Remarkably, we once also experimentally observed two Si dopants moving under electron irradiation from two sites' separation to become nearest neighbours (see Ref.~\cite{SUPPL}).

To conclude, we demonstrated how 60 keV electron irradiation causes
structural rearrangements at silicon dopant sites in the graphene lattice.
Despite appearances, the Si atoms themselves are not perturbed by the electron
beam, but undergo a random walk in the lattice due to structural relaxation taking place during a
sub-threshold electron impact on a neighbouring carbon atom. The position of this 
carbon atom thus determines the direction of the walk. Therefore, restricting intense dosing 
only to a desired carbon atom should allow the motion of the Si atoms to be 
controlled with atomic precision, and arbitrary arrangements of several Si could plausibly be attained.

\begin{acknowledgments}
We thank CSC Finland Ltd and the Vienna Scientific Cluster for extensive grants
of computational resources, and Ask Hjorth Larsen for help setting up the LCAO calculations. T.S. was supported by the Austrian Science Fund (FWF) through grant M 1497-N19,
by the Finnish Cultural Foundation, and by the Walter Ahlstr\"{o}m Foundation.
J.K. was supported by the FWF via M~1481-N20, and by the University of Helsinki Funds. J.C.M. and C.M. acknowledge support from the FWF project P25721-N20 and the European Research Council (ERC) grant PICOMAT.
SuperSTEM is the UK National Facility for Aberration-Corrected STEM, supported by the Engineering and Physical Sciences Research Council (EPSRC), who also supported R.Z. and U.B. via grant EP/I008144/1.\end{acknowledgments}

%




\setcounter{figure}{0}

\makeatletter 
\renewcommand{\thefigure}{S\@arabic\c@figure}
\makeatother

\newpage{}
\begin{center}\LARGE{Supplemental Material}\end{center}

\section{Computational unit cell}
The unit cell for the molecular dynamics calculations needs to be large enough so that atoms at the edge of the cell do not appreciably move before the ejected atom has either escaped or reached a turning point in its trajectory, as this would cause an overestimation of the restoring forces. In our case a 8$\times$6 graphene supercell of 96 atoms was sufficient. 

\section{Total energy calculations}
All calculations described below were performed in a larger 10$\times$8 unit cell of 160 atoms (+ possible adatoms) to avoid finite size effects, using a $3\times3\times1$ $\mathbf{k}$-point grid. The structures were allowed to fully relax without constraints so that the maximum forces were lower than 0.01 eV/\AA. 

\subsection{Relative stabilities}
The relative stabilities of the Si-C$_{3}$ and Si-C$_{4}$ configurations were calculated as
\begin{eqnarray}
 E_{rel}&=&E_{form}^{\mathrm{Si-C}_{3}}-E_{form}^{\mathrm{Si-C}_{4}} \nonumber \\
 &=&(E_{tot}^{\mathrm{Si-C}_{3}}+E_{C}-E_{tot}^{\mathrm{gra}}-E_{Si})-(E_{tot}^{\mathrm{Si-C}_{4}}+2E_{C}-E_{tot}^{\mathrm{gra}}-E_{Si})\\
&=& E_{tot}^{\mathrm{Si-C}_{3}}-(E_{tot}^{\mathrm{Si-C}_{4}}+E_{C}) \nonumber
\end{eqnarray}
where $E_{form}$ denote the formation energies of the two configurations, $E_{tot}$ are their fully relaxed total energies, $E_{tot}^{\mathrm{gra}}$ is the total energy of pristine graphene system, $E_{C}$ is the chemical potential for carbon in graphene (calculated at -9.2232 eV), and $E_{Si}$ is the energy of an isolated Si atom (note that this cancels out in the calculation). This results in a relative energy of -1.023 eV, establishing Si-C$_{3}$ as the more stable configuration.

\subsection{Si-C$_{3}$ flipping barrier}
\begin{figure}
  \includegraphics{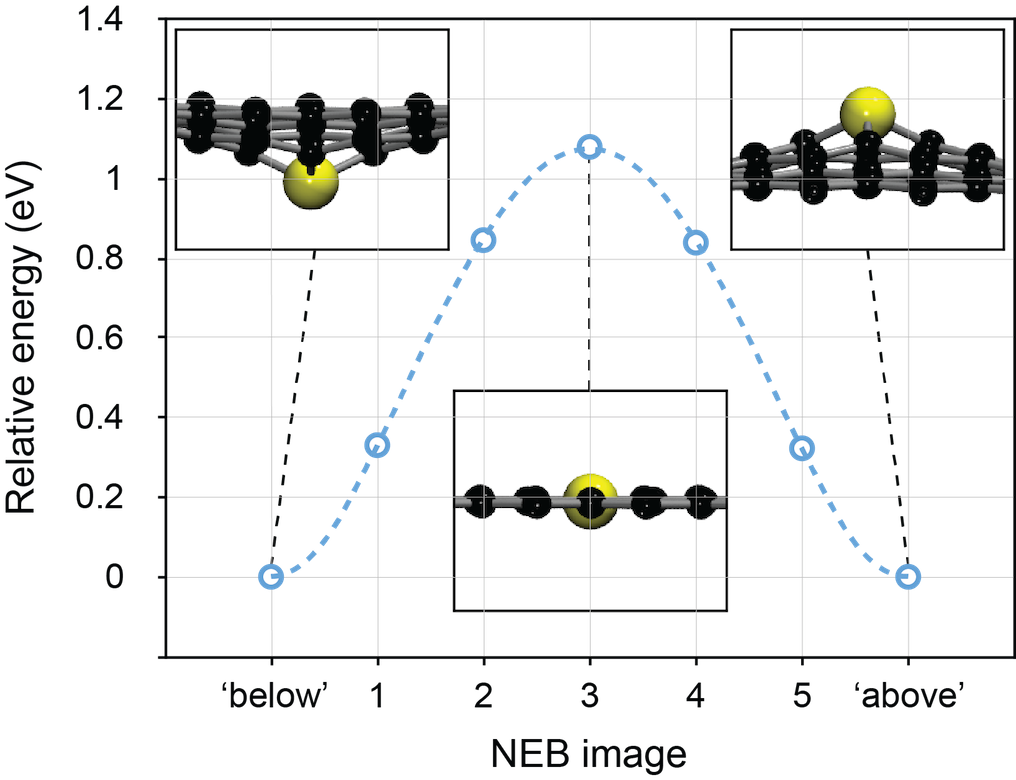}
  \caption{A nudged elastic band calculation for the energy barrier of flipping the Si-C$_{3}$ dopant from `below' to `above' the graphene plane. The dotted trendline is a fourth order polynomial fit to the energies, and the insets show the starting, midpoint and final configurations.\label{NEB}}
\end{figure}

The assess the energy barrier for flipping the Si dopant bonded in the Si-C$_{3}$ configuration from `below' to `above' the graphene plane, we performed a simple nudged elastic band calculation (G. Henkelman and H. J\'{o}nsson, Journal of Chemical Physics \textbf{113}, 9978 (2000)). Since the relaxed starting and ending configurations are symmetric with respect to the graphene plane, the trajectory is a simple line with the maximum energy at the midpoint where the Si atom is in the plane of the lattice. The energy barrier given by the calculation is 1.08 eV (Figure \ref{NEB}), and we also confirmed by a DFT/MD calculation that already a 2 eV starting kinetic energy was sufficient to dynamically induce the flip.

\subsection{Si-C$_{4}$ $\rightarrow$ Si-C$_{3}$ reconversion}
\begin{figure}[t]
  \includegraphics{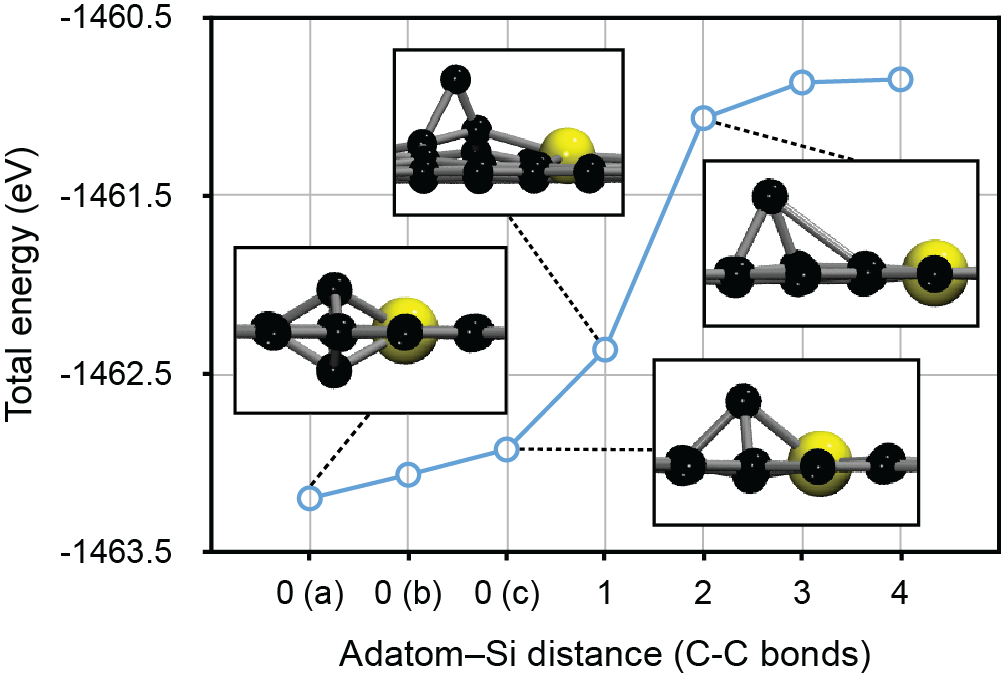}
  \caption{Total energies of the Si-C$_{4}$ system with a single carbon adatom initially placed on a C-C bridge site 1\ldots4 bonds away from the defect. During relaxation, adatoms placed 1 or 2 bonds away spontaneously relaxed closer to the defect, resulting in several configurations with the C bonded directly to the Si.\label{1SiV+C}}
\end{figure}

Since each of the Si-C$_{4}$ we observed spontaneously reconverted back into the more energetically stable Si-C$_{3}$ configuration---presumably by adatom diffusion and recombination---we studied the energetics of the Si-C$_{4}$ defect and a single C adatom (Fig. \ref{1SiV+C}). In the initial configurations, the C adatom was placed on a C-C bridge site 1\ldots4 bonds away from the defect site, and then the structures were fully relaxed. During geometry optimisation, we observed that adatoms placed 1 or 2 bonds away spontaneously moved closer to the defect site, bonding directly with the Si atom. Thus although there seems to be an energy barrier for the recombination that our geometry optimisation could not overcome, there is a clear attractive interaction between the Si-C$_{4}$ defect and any diffusing C adatoms in the system.

\newpage
\subsection{Two nearby Si atoms}

\begin{figure}[h]
  \includegraphics{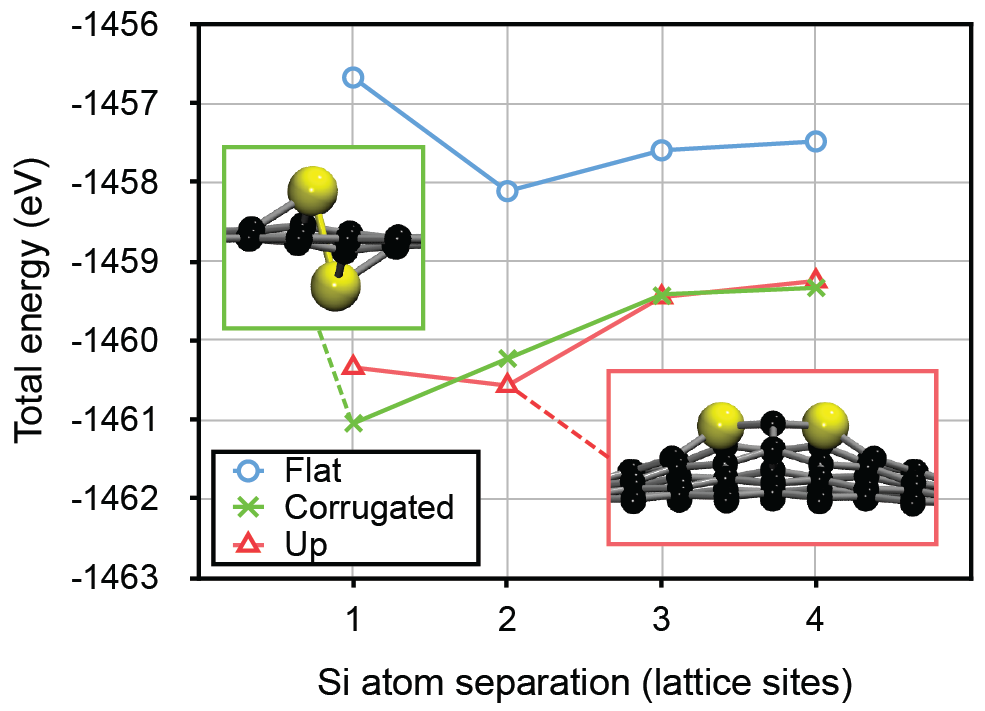}
  \caption{Total energies of systems with two silicon dopants incorporated into the lattice, separated by 1\ldots4 lattice sites. `Flat' refers to a configuration where both atoms are in the same plane as the graphene lattice (stable but energetically unfavourable), `Corrugated' to where the Si atoms protrude on different sides of the lattice, and `Up' to where both protrude unidirectionally on one side. The insets show the most stable `Corrugated' (green) and `Up' (red) configurations.\label{Si-Si}}
\end{figure}

To explore the possibility of moving several Si dopants close to each other, we calculated the energetics of systems with two Si dopants incorporated into the graphene lattice separated by 1\ldots4 lattice sites (Figure \ref{Si-Si}). We found that the energies of the configurations with both Si atoms on either one (unidirectional) or on different (corrugated) sides of the graphene lattice were almost equal, apart from the single lattice site separation, where the corrugated configuration was more stable. Furthermore, the energies were lowered when the two Si atoms were brought closer together, with a shallow energy minimum for the unidirectional configuration found at a two lattice sites' separation.

\begin{figure}
  \includegraphics{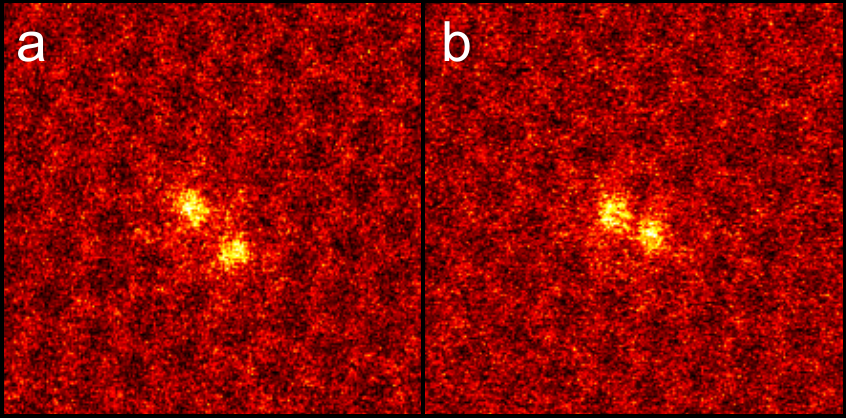}
  \caption{Two Si dopants initially separated by two lattice sites
move closer to each other under 60 keV electron irradiation (HAADF).
Each image cropped from an average of six 2.44 second frames covering a larger field of view. (a) The starting
configuration shows two bright Si atoms located as next nearest
neighbours in the lattice. (b) The same area after a dose of 1.4$\times10^{7}$ e$^{-}$/\AA$^{2}$. The lower right Si atom has moved one lattice site closer, making the two Si dopants nearest neighbours.\label{2Si2-1}}
\end{figure}

In one particular instance, we experimentally observed two Si dopants near each other in the lattice (see Figure \ref{2Si2-1}). In the beginning of observation, the Si atoms occupied next nearest neighbour lattice positions. After a certain time of continuous imaging, one of the Si atoms moved one lattice site closer, making the two Si atoms nearest neighbours in the lattice (with one carbon atom possibly removed) This offers direct experimental support for the calculations shown in Figure \ref{Si-Si}, demonstrating that at least two Si atoms can be brought arbitrarily close.

\newpage
\clearpage
\section{Statistical analyses}

\begin{figure}[b]
\includegraphics[width=0.4\linewidth]{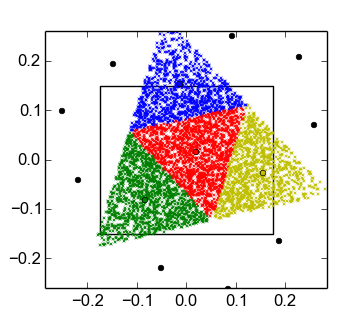}
\caption{An example Monte Carlo simulation used to estimate the precise dose on each of the C neighbours to the Si dopant for each frame. The scan frame is denoted by the inner square and the Si atom by the open circle in the centre of the triangle formed from the red dots representing the electrons impinging the Si atom. Electrons impinging on each of the neighbouring three carbon atoms are represented by the blue, green and yellow dots. The axes denote distances in nm.\label{MCfig}}
\end{figure}

The doses for the 38 jump events observed in our experimental data can be found in Table I. An illustration of the Monte Carlo integration used to estimate the precise irradiation doses on each of the carbon atoms neighbouring the Si dopant the scan frame was centred on can be seen in Figure \ref{MCfig}. If the integration resulted in a proportion of dose larger than 1 on the C atom that jumped (that is, if that C received a higher dose than the Si due to the position of the frame), the precise event dose was simply equal to the experimental dose.

Assuming the jump data are stochastic, the waiting times (or, equivalently, the doses) should arise from a Poisson process with mean $\lambda$. Thus the probability to find $k$ events in a given time interval follows the Poisson distribution

\begin{equation}
f(k;\lambda) = \textrm{Pr}(X=k)=\frac{\lambda^{k}e^{-\lambda}}{k!}.
\end{equation}

To estimate the Poisson expectation value, the cumulative doses (calculated by summing the doses incurred by those specific carbon atoms that were observed to jump) given in Table I were combined into a single dataset, which was divided into bins of width $w$, and the number of bins with 0, 1, 2\ldots occurrences were counted. 
The binwidth was optimised by minimising the Pearson $\chi^{2}$ statistic calculated between a fitted Poisson distribution and the data (see below). The mean $\lambda$ was estimated for the dataset by maximising the log-likelihood function of the fit. 

An error estimate for the mean was calculated using the approximate confidence interval proposed for Poisson processes with small means and small sample sizes by Khamkong (M. Khamkong, Open Journal of Statistics \textbf{02}, 204 (2012)): 

\begin{equation}
\textrm{CI}_{95\%} = \bar{\lambda} + \frac{Z^{2}_{2.5}}{2n}\pm Z_{2.5}\sqrt{\frac{\bar{\lambda}}{n}},
\end{equation}
where $\bar{\lambda}$ is the estimated mean and $Z_{2.5}$ is the normal distribution single tail cumulative probability corresponding to a confidence level of $(100-\alpha)=95\%$, equal to 1.96.

The statistical analyses were conducted using the Wolfram Mathematica software, and the Computable Document Format Mathematica script used is included as a Supplemental File. For the entire dataset, a binwidth of $1.06\times10^{8}$ e$^{-}$/\AA$^{2}$ was found to be optimal, yielding an estimated mean dose of $2.57\times10^{8}$ e$^{-}$/\AA$^{2}$. However, note that when the experimental cumulative dose data is plotted along with 20 simulated random Poisson processes with the fitted mean, we see that the agreement between the data and the fit is poor due to the influence of high dose events (ie., the outliers) showing as high steps in the cumulative data (see Figure S6).

We then calculated the probabilities that each event was a random event corresponding to a Poisson process with the fitted mean, also shown Table I. This reveals that the three events with the highest doses have probabilities lower than $10^{-3}$ to be a result of the same process as the rest of the cases. Reanalysis of the data without these outliers results in a revised expectation value of 1.63$\times10^{8}$~e$^{-}$/\AA$^{2}$ (optimal binwidth $1.215\times10^{8}$ e$^{-}$/\AA$^{2}$). Although the Poisson fit is seemingly slightly less good than with the full dataset, a much better fit between the experimental cumulative doses and the simulated random Poisson processes with the fitted mean can be seen (Figure S7). The expectation value was found to not be very sensitive to the value of the binwidth, provided the chosen width resulted in a good fit.

\section{Jump in grain boundary}
Figure S8 shows a Si atom moving by one lattice  site along a graphene grain boundary under electron irradiation.

\begin{figure}
\includegraphics[width=1.0\linewidth]{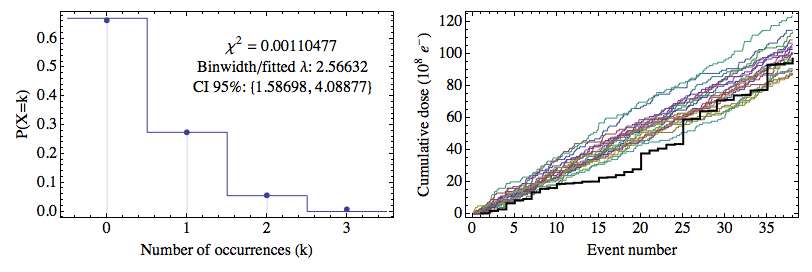}
\caption{Poisson analysis of the entire dataset.\label{fulldata}}
\end{figure}

\begin{figure}[h]
\includegraphics[width=1.0\linewidth]{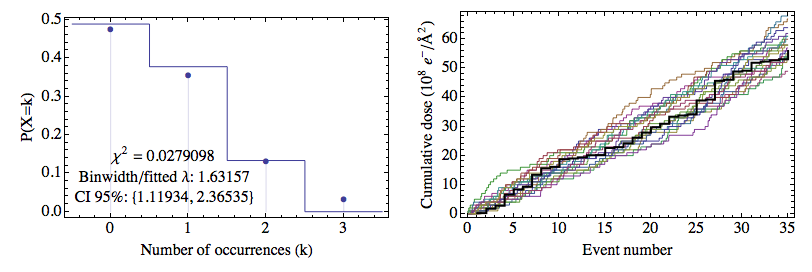}
\caption{Poisson analysis of the dataset with the outliers excluded.\label{excldata}}
\end{figure}

\begin{figure}[t]
\includegraphics[]{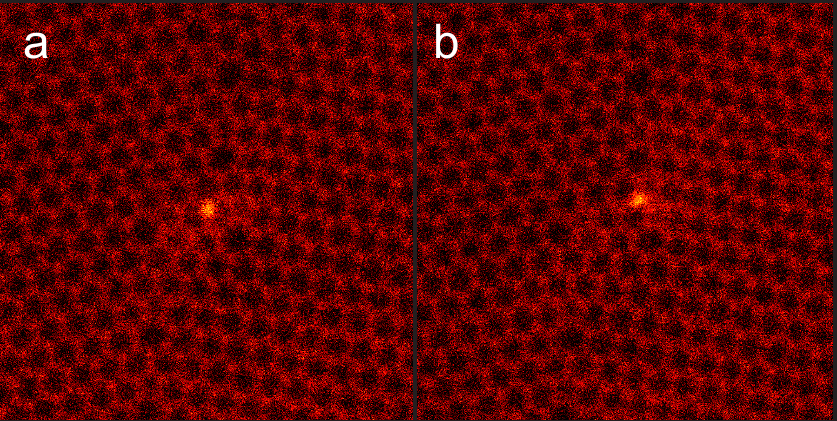}
\caption{(Color online) Si atom moving in a graphene grain boundary under
60~keV electron irradiation (MAADF detector). (a) The starting configuration,
with the Si atom visible by its brighter Z contrast. (b) The same area imaged
after the Si atom had moved by one lattice site along the grain
boundary.\label{FigGB}}
\end{figure}

\begin{table}[h]
\caption{\label{datatable}Experimental data of the Si jump events. The columns show: binned electron micrograph frame number (0.5 to 0.94 s per frame depending on the experiment); the dose for each event (waiting time multiplied by dose rate); cumulative dose from the start of each experiment; proportion of dose on carbon neighbour number one (C\#1; proportion on the Si was by definition 1) estimated by the Monte Carlo calculation; proportion on C\#2; proportion on C\#3; number of the C atom towards which the Si was observed to jump; precise dose on the C atom that was observed to jump (note that the maximum proportion was set to 1); precise cumulative dose on that C atom; and the probability for this dose assuming a Poisson process with the fitted mean. *Dose per area, in units of 10$^{8}$ e$^{-}$/\AA$^{2}$.}
\begin{tabular}{p{1.3cm}|p{1.2cm}|p{1.2cm}|p{1.2cm}|p{1.2cm}|p{1.2cm}|p{1.5cm}|p{1.5cm}|p{1.5cm}|p{1.7cm}}
\textbf{Frame \#} & \textbf{Event dose*} & \textbf{Cum. dose*} & \textbf{Prop. on C \#1} & \textbf{Prop. on C\#2} & \textbf{Prop. on C\#3} & \textbf{\#C that jumped} & \textbf{Precise dose on C*} & \textbf{Precise cum. dose*} & \textbf{Poisson probability}\\
\hline
4 & 0.47 & 0.47 & 1.004 & 0.496 & 0.404 & 2 & 0.23 & 0.23 & 0.923 \\
24 & 2.35 & 2.81 & 0.299 & 0.963 & 0.641 & 3 & 1.50 & 1.74 & 0.726 \\
42 & 2.11 & 4.93 & 0.691 & 0.488 & 0.733 & 2 & 1.03 & 2.77 & 0.726 \\
84 & 4.93 & 9.85 & 0.402 & 0.775 & 0.675 & 2 & 3.79 & 6.56 & 0.257 \\
120 & 2.81 & 12.66 & 0.691 & 0.562 & 0.652 & 3 & 1.83 & 8.39 & 0.726 \\
129 & 1.06 & 13.72 & 0.423 & 0.546 & 0.886 & 3 & 0.94 & 9.33 & 0.923 \\
182 & 6.22 & 19.94 & 0.642 & 0.661 & 0.595 & 2 & 4.10 & 13.43 & 0.118 \\
203 & 2.46 & 22.40 & 0.376 & 0.572 & 0.895 & 3 & 2.19 & 15.62 & 0.473 \\
210 & 0.82 & 23.22 & 0.626 & 0.491 & 0.778 & 1 & 0.52 & 16.14 & 0.923 \\
247 & 4.34 & 27.56 & 0.647 & 0.603 & 0.653 & 2 & 2.60 & 18.74 & 0.473 \\
252 & 0.59 & 28.14 & 0.549 & 0.852 & 0.473 & 3 & 0.28 & 19.02 & 0.923 \\
257 & 0.47 & 28.61 & 0.504 & 0.574 & 0.836 & 3 & 0.39 & 19.41 & 0.923 \\
268 & 1.06 & 29.67 & 0.586 & 0.612 & 0.679 & 3 & 0.72 & 20.13 & 0.923 \\
270 & 0.23 & 29.90 & 0.391 & 0.268 & 1.364 & 2 & 0.06 & 20.20 & 0.923 \\
304 & 3.99 & 33.89 & 0.545 & 0.732 & 0.607 & 3 & 2.43 & 22.63 & 0.473 \\
309 & 0.59 & 34.48 & 0.469 & 0.486 & 0.917 & 1 & 0.28 & 22.90 & 0.923 \\
324 & 2.23 & 36.70 & 0.660 & 0.633 & 0.601 & 3 & 1.34 & 24.24 & 0.726 \\
347 & 2.70 & 39.40 & 0.653 & 0.520 & 0.725 & 3 & 1.94 & 26.18 & 0.726 \\
360 & 2.70 & 42.10 & 0.490 & 0.644 & 0.754 & 2 & 1.73 & 27.91 & 0.726 \\
\hline
\end{tabular}
\end{table}

\begin{table}[h]
\begin{tabular}{p{1.2cm}|p{1.2cm}|p{1.2cm}|p{1.2cm}|p{1.2cm}|p{1.2cm}|p{1.5cm}|p{1.5cm}|p{1.5cm}|p{1.7cm}}
\textbf{Frame \#} & \textbf{Event dose*} & \textbf{Cum. dose*} & \textbf{Prop. on C \#1} & \textbf{Prop. on C\#2} & \textbf{Prop. on C\#3} & \textbf{\#C that jumped} & \textbf{Precise dose on C**} & \textbf{Precise cum. dose**} & \textbf{Poisson probability}\\
\hline
27 & 10.94 & 53.04 & 0.906 & 0.913 & 0.957 & 1 & 9.91 & 37.82 & 3.40$\times10^{-4}$ \\
32 & 2.03 & 55.07 & 0.903 & 1.028 & 1.056 & 1 & 1.83 & 39.65 & 0.726 \\
35 & 1.22 & 56.29 & 1.040 & 1.099 & 1.045 & 3 & 1.22 & 40.87 & 0.726 \\
44 & 3.65 & 59.93 & 0.963 & 0.741 & 0.667 & 3 & 2.43 & 43.30 & 0.473 \\
45 & 0.41 & 60.34 & 0.973 & 0.994 & 1.046 & 3 & 0.41 & 43.70 & 0.923 \\
84 & 15.40 & 75.74 & 2.500 & 1.429 & 1.286 & 1 & 15.40 & 59.11 &  1.53$\times10^{-8}$ \\
86 & 0.41 & 76.15 & 1.086 & 1.029 & 1.040 & 1 & 0.41 & 59.51 & 0.923 \\
97 & 5.27 & 81.42 & 1.003 & 0.952 & 0.901 & 3 & 4.75 & 64.26 & 0.118 \\
\hline
191 & 0.41 & 81.82 & 0.800 & 0.920 & 0.840 & 3 & 0.34 & 64.60 & 0.923 \\
207 & 6.48 & 88.30 & 1.013 & 0.961 & 0.992 & 3 & 6.42 & 71.02 & 0.016 \\
208 & 0.41 & 88.71 & 0.966 & 0.483 & 0.862 & 1 & 0.39 & 71.41 & 0.923 \\
215 & 2.84 & 91.54 & 0.993 & 0.920 & 1.014 & 1 & 2.81 & 74.22 & 0.473 \\
216 & 0.41 & 91.95 & 0.750 & 0.594 & 0.688 & 1 & 0.30 & 74.52 & 0.923 \\
\hline
295 & 2.84 & 94.79 & 0.960 & 0.977 & 1.080 & 1 & 2.72 & 77.25 & 0.473 \\
296 & 0.41 & 95.19 & 1.083 & 0.667 & 1.333 & 1 & 0.41 & 77.65 & 0.923 \\
335 & 15.81 & 111.00 & 1.034 & 1.034 & 1.051 & 1 & 15.81 & 93.46 &  1.53$\times10^{-8}$ \\
\hline
410 & 0.41 & 111.41 & 1.263 & 0.842 & 1.316 & 3 & 0.41 & 93.87 & 0.923 \\
411 & 0.41 & 111.81 & 1.042 & 1.042 & 1.208 & 3 & 0.41 & 94.27 & 0.923 \\
418 & 2.84 & 114.65 & 0.902 & 1.049 & 1.049 & 3 & 2.84 & 97.11 & 0.473 \\
\end{tabular}
\end{table}



\begin{thebibliography}{36}
\bibitem [{not({\natexlab{a}})}]{note1}%
  \BibitemOpen
  \href@noop \bibinfo {The FD mode should in
  principle give more accurate forces than the LCAO mode, but the inherent
  inaccuracy in the numerical integration of the equations of motion in the MD
  algorithm may render this irrelevant for the actual simulated trajectories.
  Thus the different values could also be interpreted as an estimate for the
  inaccuracy of the calculations.}\BibitemShut {Stop}%
\bibitem [{not({\natexlab{b}})}]{note2}%
  \BibitemOpen
  \href@noop \bibinfo {Our average dose per
  event was 2.55$\times 10^{8}$~e$^{-}$/\r A$^{2}$, geometrical mean
  1.22$\times 10^{8}$~e$^{-}$/\r A$^{2}$. These would have resulted in
  estimated cross sections of 0.392 and 0.820~barn.}\BibitemShut {Stop}%
\end{thebibliography}
\end{document}